\documentclass[10pt,nocopyrightspace]{sigplanconf}

\usepackage{amsmath}
\usepackage[dvips]{graphicx}
\graphicspath{{./figures/}}
\usepackage[caption=false,font=footnotesize]{subfig}
\usepackage{listings}
\usepackage{verbatim}
\usepackage{url}
\usepackage{multirow}

\usepackage{footnote}

\usepackage{hyperref}
\hypersetup{
    pdftitle={Auto-tuning of Local Memory Use on GPGPUs},
    pdfauthor={Tianyi David Han, Tarek S. Abdelrahman},
}

\usepackage{flushend}

\newcommand{\code}[1]{\texttt{#1}}

\newcommand{\wi}{workitem}
\newcommand{\wis}{workitems}
\newcommand{\wg}{workgroup}
\newcommand{\wgs}{workgroups}

\begin{document}

\titlebanner{}
\preprintfooter{}

\title{Automatic Tuning of Local Memory Use on GPGPUs}

\authorinfo{Tianyi David Han\and Tarek S.\ Abdelrahman}
{The Edward S.\ Rogers Sr.\ Department of
    Electrical and Computer Engineering\\
        University of Toronto\\
        Toronto, Ontario M5S 3G4, Canada}
{\{han,tsa\}@eecg.toronto.edu
\vspace*{-0.1in}
}

\maketitle


\begin{abstract}

The use of local memory is important to improve the performance of OpenCL
programs.
However, its use may not always benefit performance, depending on various
application characteristics, and there is no simple heuristic for deciding
when to use it.
We develop a machine learning model to decide if the optimization is
beneficial or not.
We train the model with millions of synthetic benchmarks and show that it can
predict if the optimization should be applied for a single array, in both
synthetic and real benchmarks, with high accuracy.

\end{abstract}


\section{Introduction}
\label{sec:intro}

The last few years have seen considerable growth in the use of Graphics
Processing Units (GPUs) to accelerate general purpose applications.
Today, the OpenCL standard~\cite{opencl} is used to express applications in
which computationally intensive segments, or \emph{kernels} are launched for
execution on the GPU.
However, to realize the performance benefits of GPUs, programmers must
\emph{optimize} their kernels to better exploit the underlying GPU
architecture~\cite{cuda-man}.


One important optimization is the use of local memory -- a small user-managed
on-chip storage, which can improve performance by an order of magnitude.
However, the optimization is not always beneficial, depending on
1) the amount of data reuse and the degree of memory non-coalescing in the
   kernel, and
2) the instruction overhead and the amount of drop in parallelism the
   optimization introduces.
The extent to which these factors influence the optimization's benefit is
often not clear.
There is no simple heuristic for deciding whether or not the optimization
should be applied.


We explore the use of machine learning to auto-tune the local memory
optimization.
We build a model that predicts the benefit of caching a region of an array in
local memory, based on the performance of a set of training kernels with and
without the optimization.
We then apply this model to a new kernel to decide if the optimization should
be applied.


A unique aspect of our work is the use of many synthetically generated kernels
for model training.
We believe that machine learning, particularly on a high-dimensional feature
space, demands a large training set which is difficult to assemble from
real-world benchmarks.
The use of synthetic benchmarks allows us to build a robust model fully
exploiting the power of machine learning.
These synthetic kernels capture common data access patterns in the domains of
dense linear algebra and structured grids.


Performance data of a large number of synthetic kernel instances (with and
without using local memory) on an NVIDIA GPU shows that the optimization
brings a wide range of kernel speedup (from $0.03\times$ to $49.6\times$).
A {\em Random Forest}~\cite{randomforest} model trained on a random 10\% of
the data can predict if the optimization is beneficial, on the remaining data,
with nearly 95\% accuracy.
This model also achieves an average of nearly 95\% accuracy on eight
real-world kernels.


To our best knowledge, we believe that this is the first work that: 1) builds
an accurate model using machine learning to auto-tune the local memory
optimization, and 2) uses a large number of synthetic benchmarks for model
training.

\begin{figure*}[ht!]
\centering
\subfloat[Synthetic kernels.]
{
\includegraphics[width=2.1in]{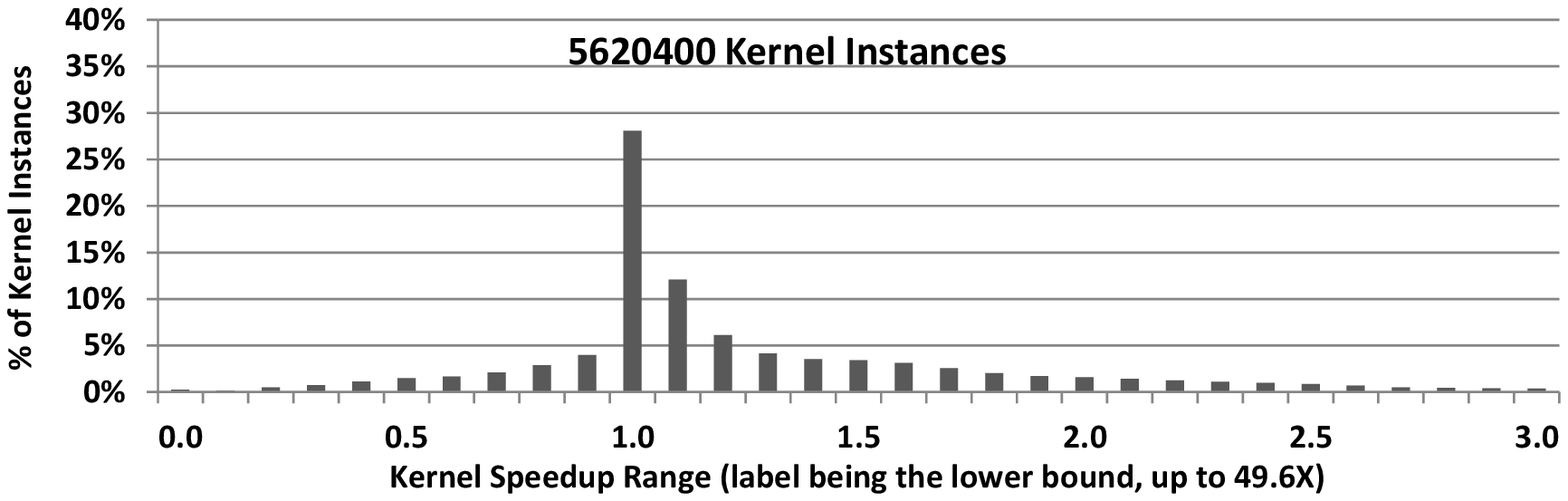}
\label{fig:speedup-hist-syn}
}
\quad
\subfloat[transpose.]
{
\includegraphics[width=2.1in]{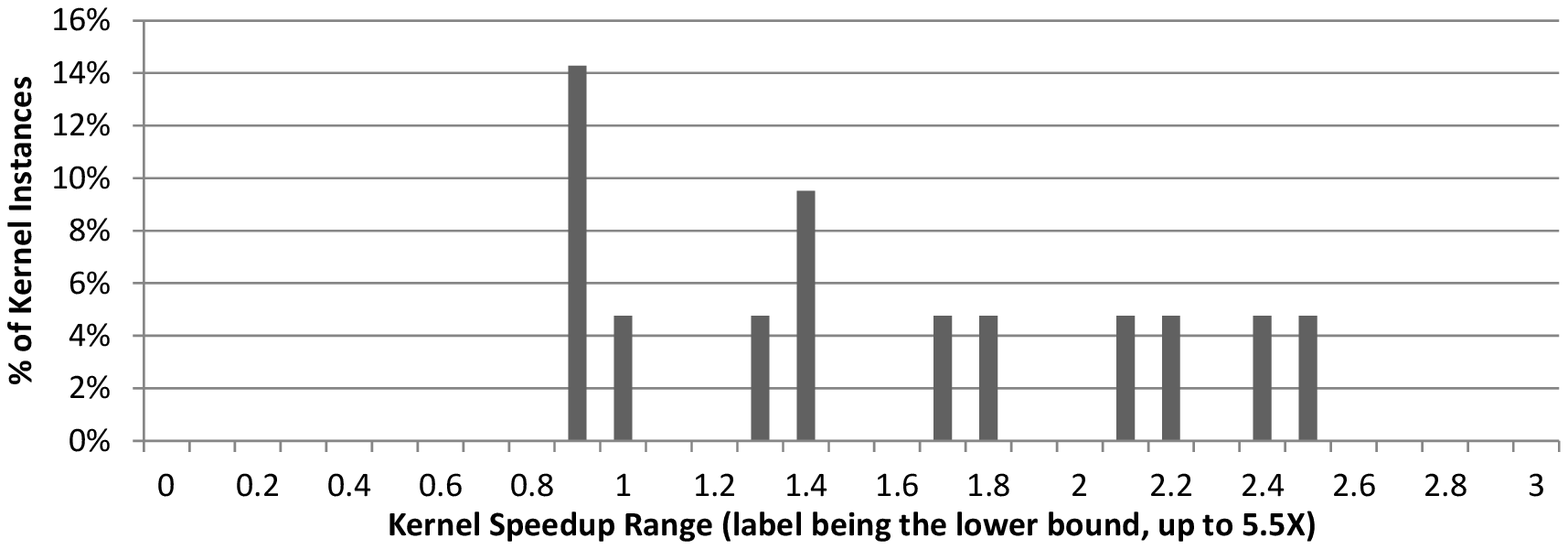}
\label{fig:speedup-hist-transpose}
}
\quad
\subfloat[matrixMul.]
{
\includegraphics[width=2.1in]{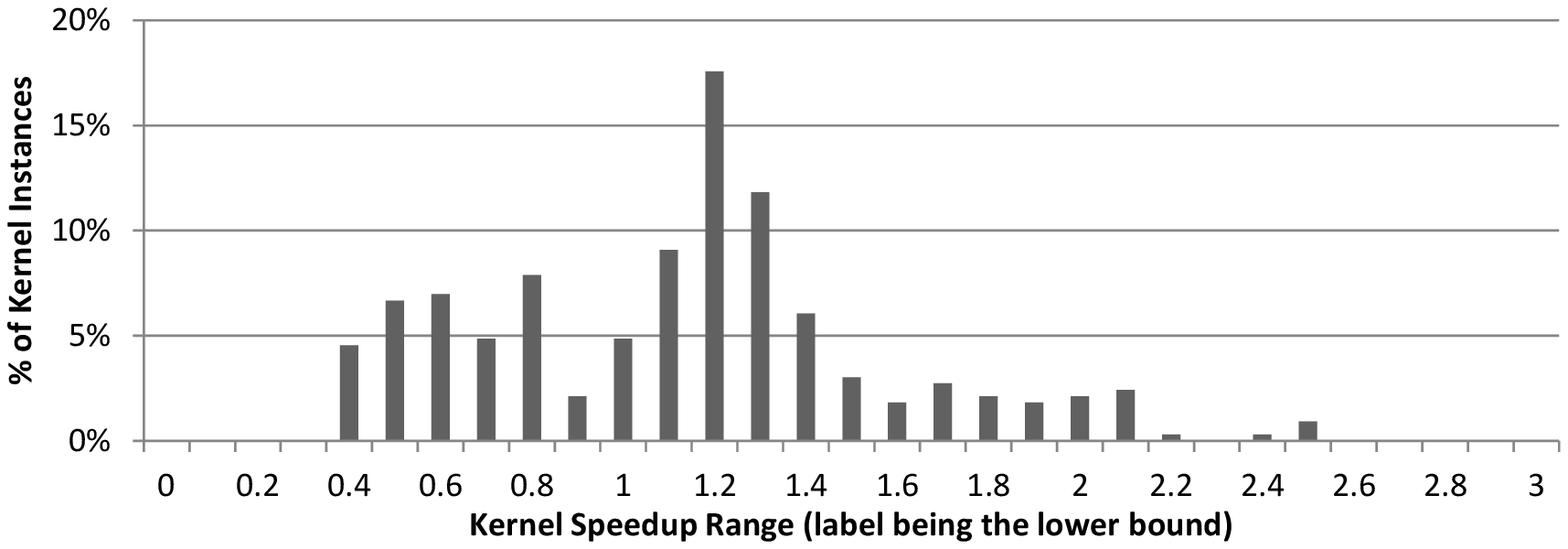}
\label{fig:speedup-hist-matrixmul}
}
\\
\subfloat[convolution.]
{
\includegraphics[width=2.1in]{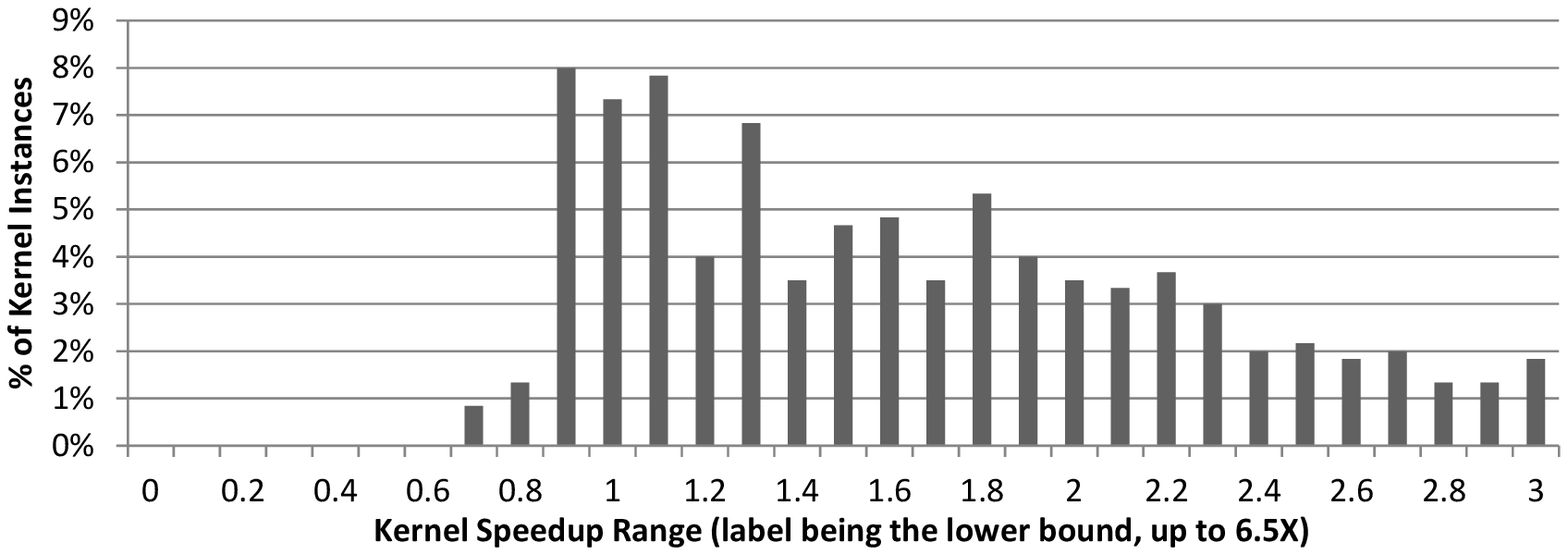}
\label{fig:speedup-hist-convsep}
}
\quad
\subfloat[MVT.]
{
\includegraphics[width=2.1in]{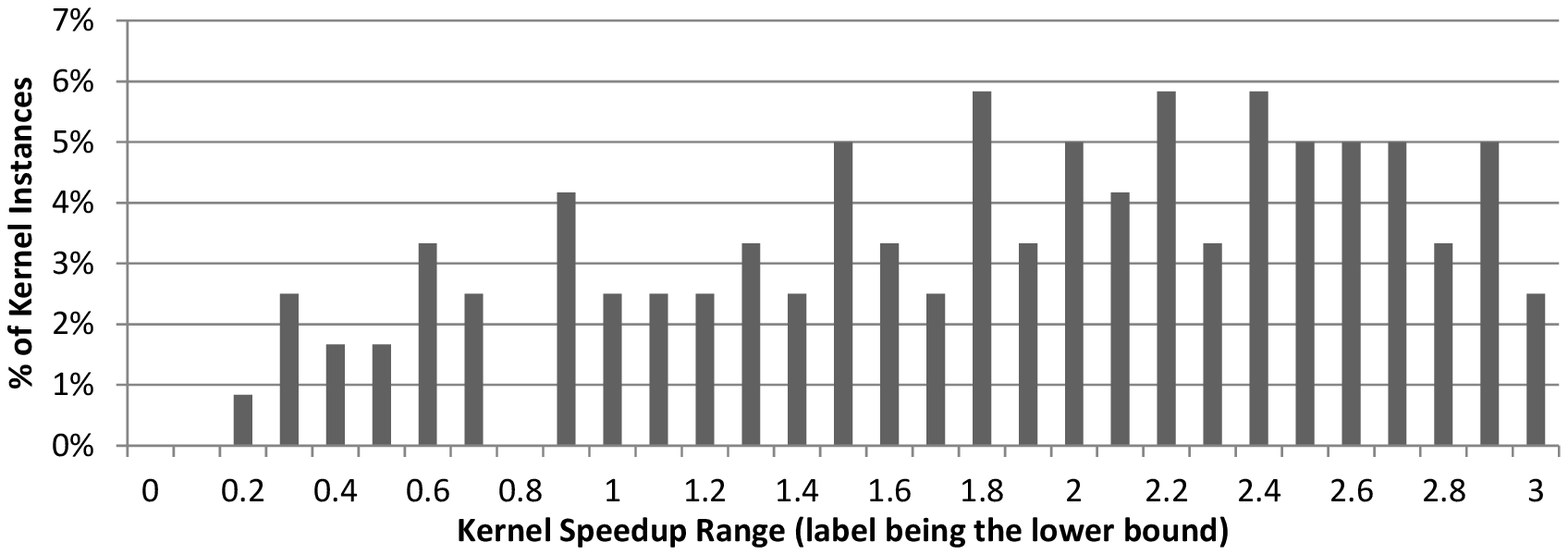}
\label{fig:speedup-hist-mvt}
}
\quad
\subfloat[SGEMM.]
{
\includegraphics[width=2.1in]{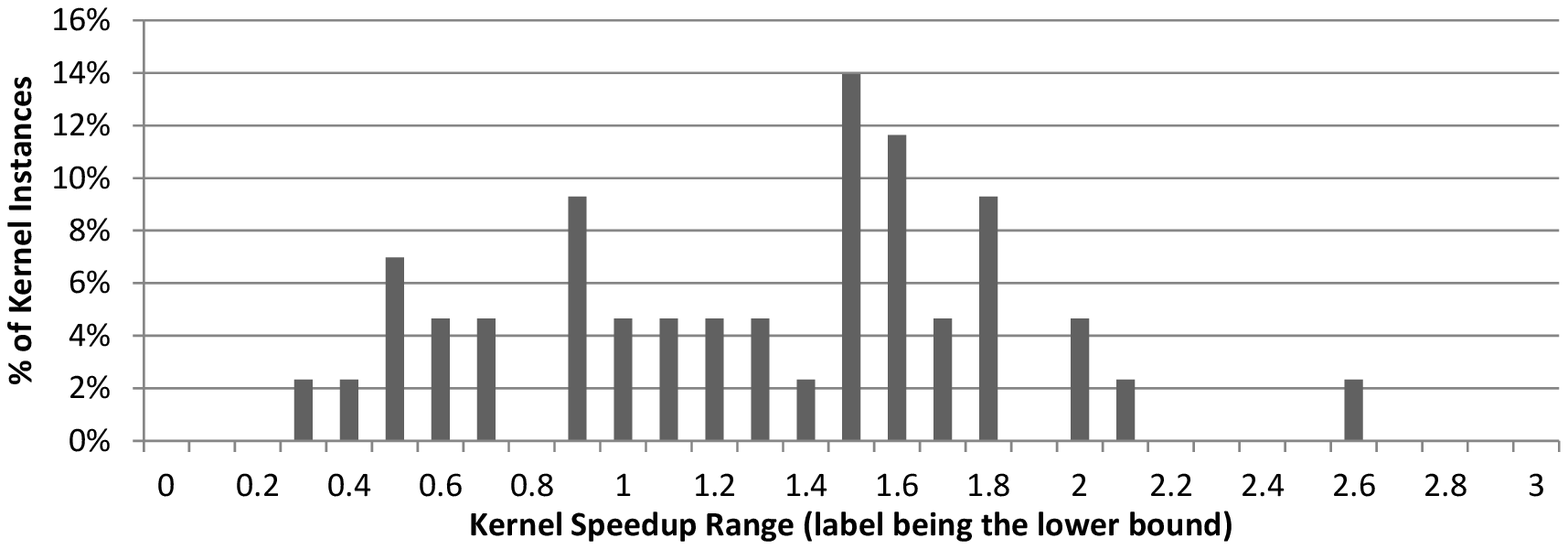}
\label{fig:speedup-hist-sgemm}
}
\\
\subfloat[SAD.]
{
\includegraphics[width=2.1in]{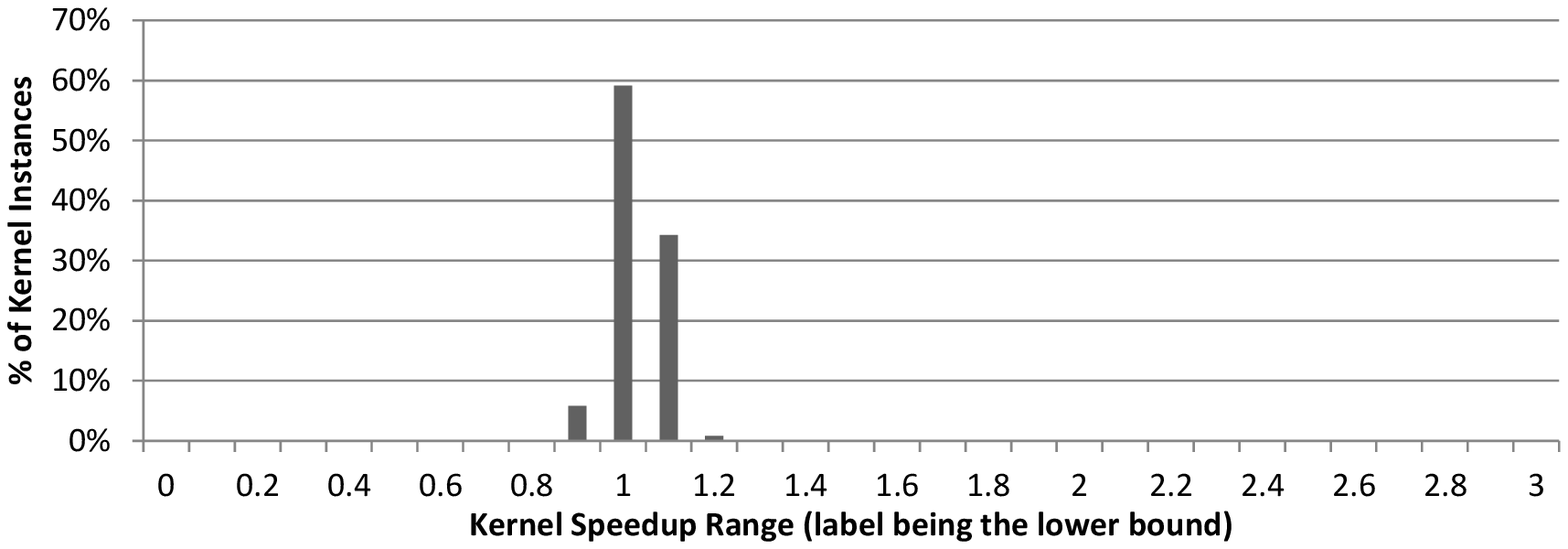}
\label{fig:speedup-hist-sad}
}
\quad
\subfloat[TPACF.]
{
\includegraphics[width=2.1in]{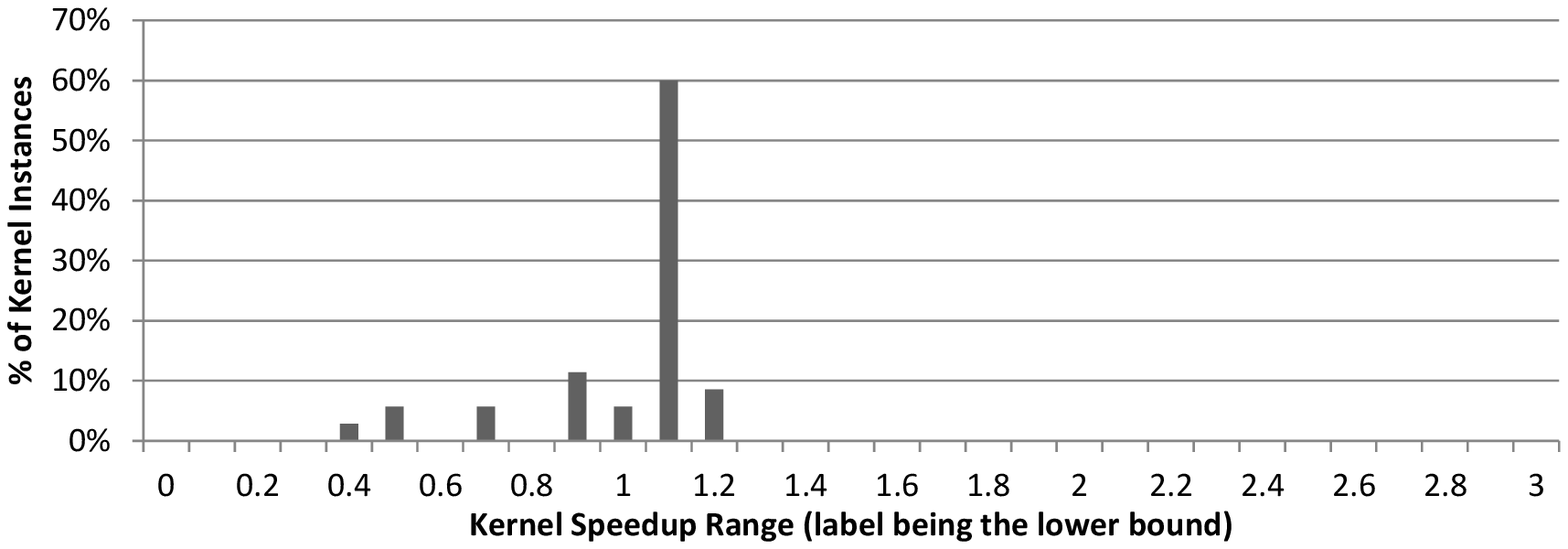}
\label{fig:speedup-hist-tpacf}
}
\quad
\subfloat[MRI-GRIDDING.]
{
\includegraphics[width=2.1in]{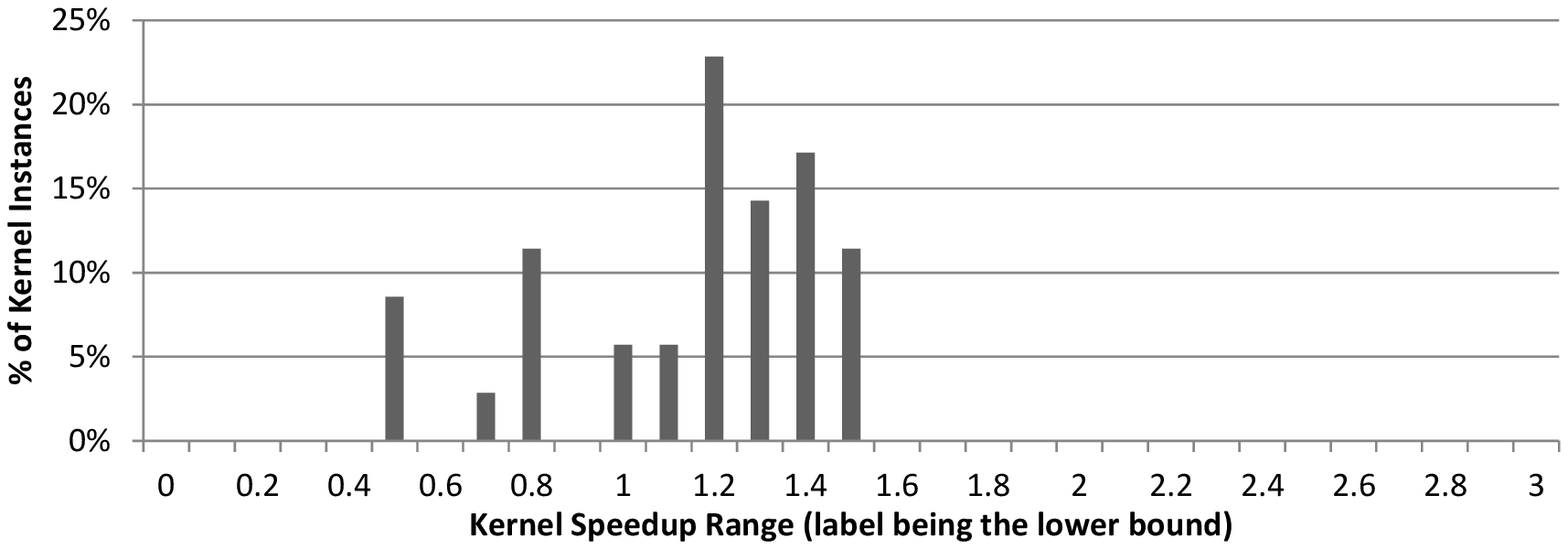}
\label{fig:speedup-hist-mri-gridding}
}
\caption{Histograms of kernel speedup brought by the local memory
optimization.}
\label{fig:speedup-histograms}
\end{figure*}


\section{Background}
\label{sec:background}


Local memory is a software-managed cache shared by all \wis\ within a \wg.
Despite being small (in kilo-bytes), its use can significantly improve kernel
performance, because it has close-to-register access latency~\cite{cuda-man}.


Local memory improves performance for two reasons.
First, caching data in local memory exploits data locality in the kernel and
can reduce the number of transactions reaching the GPU DRAM.
Data locality in a kernel (executed by many threads) can be classified as {\em
temporal} or {\em spatial}, and {\em intra-thread} or {\em inter-thread}.
Local memory is commonly used to exploit all four categories of data locality
except intra-thread temporal locality,
where the GPU compilers are able to put the data in thread-private registers.
The benefit of using local memory increases with the amount of data reuse.


Even in the absence of data reuse, local memory can improve kernel performance
by transforming non-coalesced memory accesses into coalesced
ones~\cite{ueng-lcpc08}.
A common scenario is when each \wi\ performs some sort of row-wise reduction,
forcing \wis\ to access a {\em column} of a 2D array at the same time.
This results in totally non-coalesced accesses that come with a high
performance penalty.
These non-coalesced accesses can be eliminated by first copying a batch of the
columns to the local memory in a coalesced manner.
Workitems access data from the local memory, still one column at a time, but
with no performance penalty.
This is because, unlike global memory, local memory does not suffer from
non-coalescing.


The copying of an array region from global memory to local memory is performed
{\em cooperatively} by all \wis\ in a \wg.
The region is divided into a sequence of row segments, each having a width of
a single DRAM transaction and is aligned to the transaction
boundary~\cite{cuda-man}.
These segments are cyclically distributed among {\em warps} in the \wg.
Elements in each segment are accessed by \wis\ in the designated warp in a
fully coalesced manner, resulting in a single DRAM transaction.
Overall, all global memory accesses made during the copying process are fully
coalesced.


\section{Impact of Using Local Memory}
\label{sec:opt-perf-impact}

While the use of local memory reduces the number of DRAM transactions, it may
not always improve the performance of the kernel as a whole.
First, the optimization introduces the overhead of copying array regions from
the global memory to the local memory.
Second, it may reduce the level of parallelism, i.e., the number of threads
that can concurrently execute on a GPU multi-processor, due to additional
resource usage.
A reduction in parallelism can hurt kernel performance in a holistic manner,
potentially exposing latencies of {\em all} memory accesses.
The extent to which this happens is a function of how many memory accesses
exist in the kernel as well as the amount of computation in the kernel that
may help {\em hide} this performance penalty.


Therefore, the performance impact of using local memory depends on various
kernel characteristics:

\begin{itemize}

\item Amount of data reuse of the array region copied to the local memory

\item Degree of memory non-coalescing of array accesses

\item Usage of registers and local memory by the optimization, which can
reduce parallelism

\item Memory accesses and computation in the unoptimized kernel, which
influences the performance impact of any parallelism drop

\end{itemize}


In order to assess the performance impact of these factors,
we synthetically generate a large number of kernels with varying values of the
characteristics listed above, and evaluated eight real-world benchmarks with
varying launch configurations and other kernel parameters
(Section~\ref{sec:eval}).
For each kernel, we empirically determine the kernel speedup of the local
memory optimization, as the ratio of the execution time of the original kernel
over that of the optimized kernel.
Figure~\ref{fig:speedup-histograms} shows the histograms of the resulting
speedup values for both synthetic and real-world kernels.
It confirms that the use of local memory is not always beneficial and its
performance impact is non-trivial to determine.


\section{ML-based Auto-tuning Framework}
\label{sec:framework}

Given a kernel (with its launch configuration) containing accesses to an array
that may exhibit data reuse or memory non-coalescing,
our framework decides if the local memory optimization would improve kernel
performance,
by caching the smallest array region that covers these accesses in the local
memory.


\begin{figure}[!ht]
\centering
\includegraphics[width=3.2in]{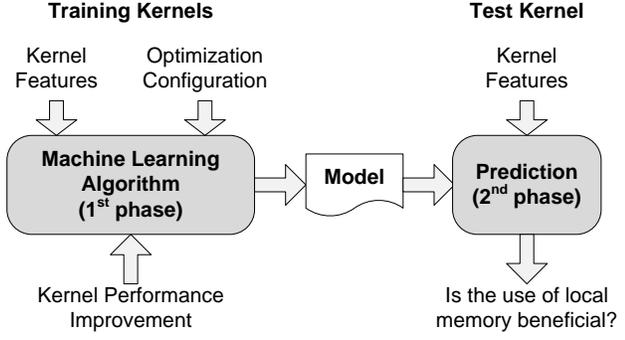}
\caption{Overview of the machine-learning-based auto-tuning framework for the
use of local memory.}
\label{fig:mlat-framework-overview}
\end{figure}


The framework consists of two phases, as shown in
Figure~\ref{fig:mlat-framework-overview}.
In the first phase (the left part of the figure),
a model is built using a machine learning algorithm, based on performance data
of a set of {\em training OpenCL kernels} with and without the local memory
optimization.
The model correlates
1) characteristics (or {\em features}) extracted from a kernel, e.g., the
   degree of data reuse and the presence of memory non-coalescing and
2) optimization configuration, i.e., the shape of the array region to be
   cache, with
3) the benefit of the optimization, e.g., kernel speedup.


In the second phase (the right part of
Figure~\ref{fig:mlat-framework-overview}),
given a new (non-training) kernel along with the candidate array accesses, we
extract kernel features as required by the model, determine the shape of the
array region to be cached in local memory, and apply the model to predict if
the optimization is beneficial.


Since a large number of training kernels is required for machine learning to
work well, we opt to synthetically generate them instead of using real-world
kernels.
In the rest of the section, we discuss the design of the synthetic kernels and
the machine learning model.


\subsection{Synthetic Benchmarks}
\label{sec:syn-bench}

We design the synthetic kernels in the form of a single kernel template with a
number of compile-time and run-time parameters.
These parameters are ``knobs'' to alter kernel characteristics that may
influence the benefit of using local memory
(Section~\ref{sec:opt-perf-impact}).


The kernel template is shown in Figure~\ref{fig:syn-kernel-template}.
It processes data from a 2D input array \code{in} and writes the result to a
2D output array \code{out}.
We call the amount of work that produces an output array element a {\em work
unit}, shown between lines 14 and 33.
It contains two nested loops followed by a code segment, which we call {\em
epilogue}.
The loop nest encloses one or more accesses to array \code{in}, interleaved
with computation (fused-multiply-add operations) and accesses to an auxiliary
input array \code{in2}, a third kernel parameter.
The epilogue also contains computation and accesses to \code{in2}, and ends
with a write to array \code{out}.
Accesses to array \code{in} are those to be possibly cached in local memory,
so we also refer to array \code{in} as the {\em target array}.
 

\begin{figure}[ht!]
\newsavebox{\sktbox}
\begin{lrbox}{\sktbox}
\begin{minipage}[b]{3.1in}
\lstset{language=c, basicstyle=\footnotesize, frame=single, numbers=left}
\begin{lstlisting}
__kernel void kmain(
    __global float *in,
    __global float *out,
    __global float *in2,
    __local float *lmem)
{
  int wg_x = get_group_id(0);
  int wg_y = get_group_id(1);
  int wi_x = get_local_id(0);
  int wi_y = get_local_id(1);
  ...
  for iter_x = 0..(NUM_WUS_X-1) {
    for iter_y = 0..(NUM_WUS_Y-1) {
      int wu_x, wu_y;
      (wu_x, wu_y) = func(wg_x, wg_y,
                          wi_x, wi_y,
                          iter_x, iter_y);
      // This is where to cooperatively load
      // a region of <in> to the local memory.
      // barrier(...);
      for i = 0..(N-1) {
        for j = 0..(M-1) {
          int idx_o = fo(wu_x, wu_y, i, j);
          int idx_i = fi(wu_x, wu_y, i, j);
          ... = in[idx_o + CO_1][idx_i + CI_1];
          ... // context (inner loop body)
          ... = in[idx_o + CO_k][idx_i + CI_k];
          ... // context (inner loop body)
        }
      }
      // barrier(...);
      ... // context (epilogue)
      out[...] = ...;
    }
  }
}
\end{lstlisting}
\end{minipage}
\end{lrbox}
\centering
\usebox{\sktbox}
\caption{Synthetic kernel template.}
\label{fig:syn-kernel-template}
\end{figure}


\begin{figure*}[ht!]
\centering
\subfloat[{\em xy-reuse}]{
\includegraphics[scale=0.4]{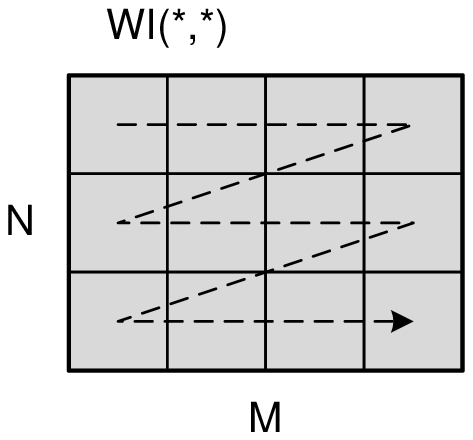}
\label{fig:hap-xy-reuse}
}
\subfloat[{\em x-reuse-row}]{
\includegraphics[scale=0.4]{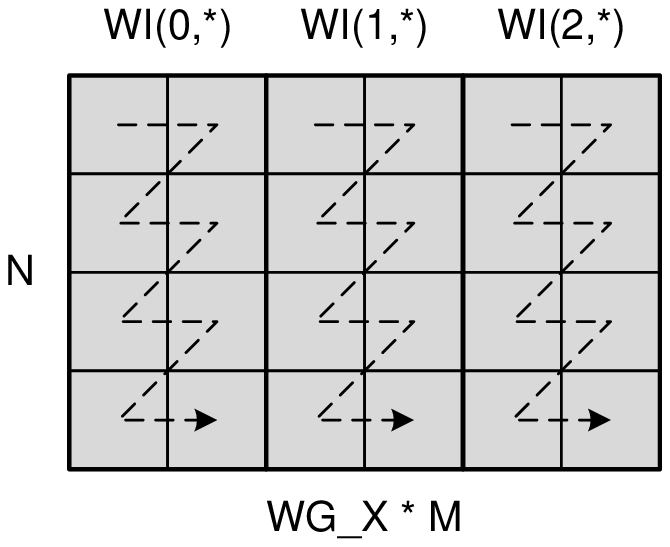}
\label{fig:hap-x-reuse-row}
}
\subfloat[{\em x-reuse-col}]{
\includegraphics[scale=0.4]{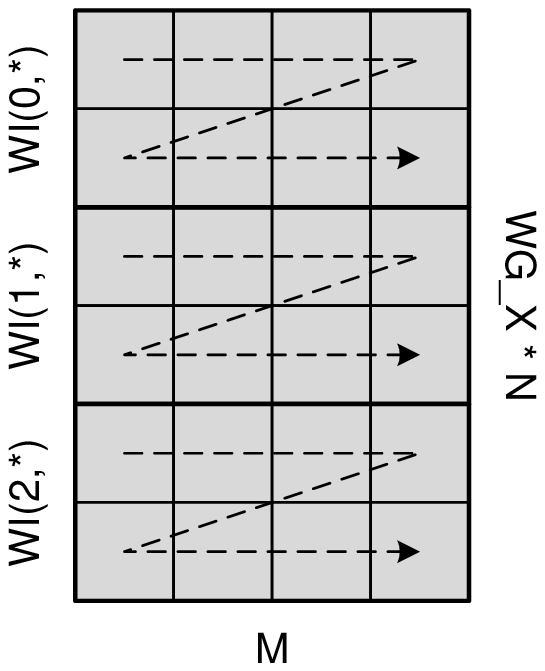}
\label{fig:hap-x-reuse-col}
}
\subfloat[{\em y-reuse-row}]{
\includegraphics[scale=0.4]{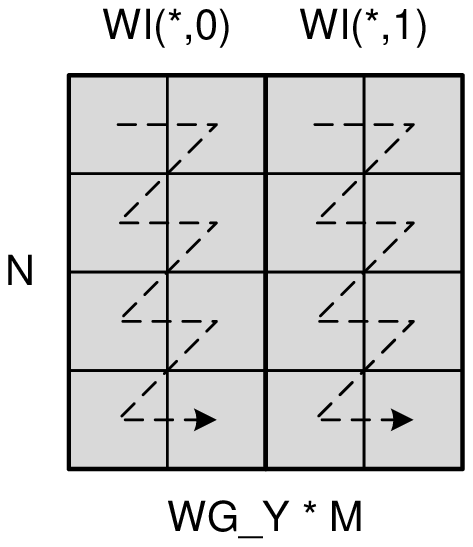}
\label{fig:hap-y-reuse-row}
}
\subfloat[{\em y-reuse-col}]{
\includegraphics[scale=0.4]{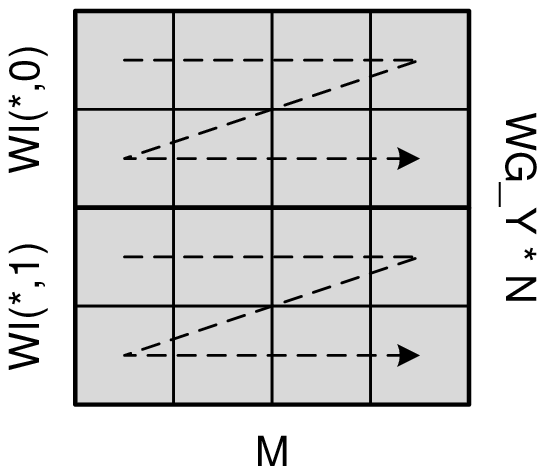}
\label{fig:hap-y-reuse-col}
}
\subfloat[{\em no-reuse-row-major}]{
\includegraphics[scale=0.4]{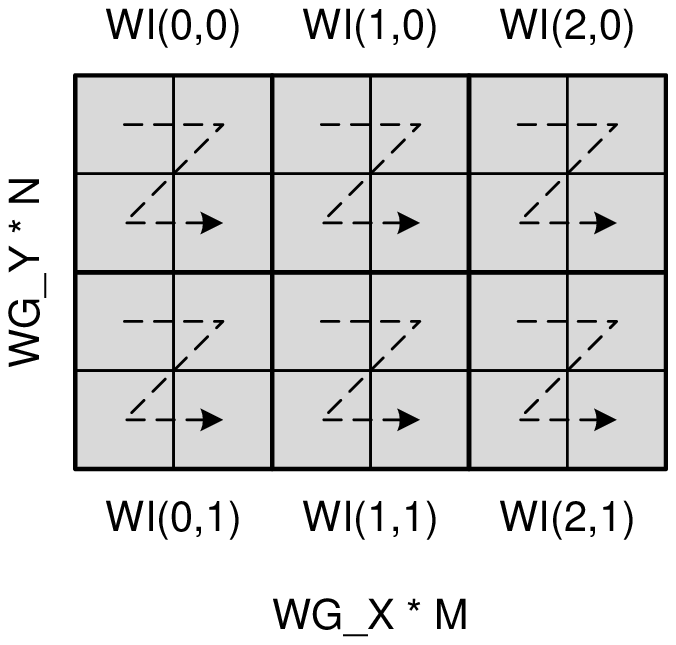}
\label{fig:hap-no-reuse-row-major}
}
\subfloat[{\em no-reuse-col-major}]{
\includegraphics[scale=0.4]{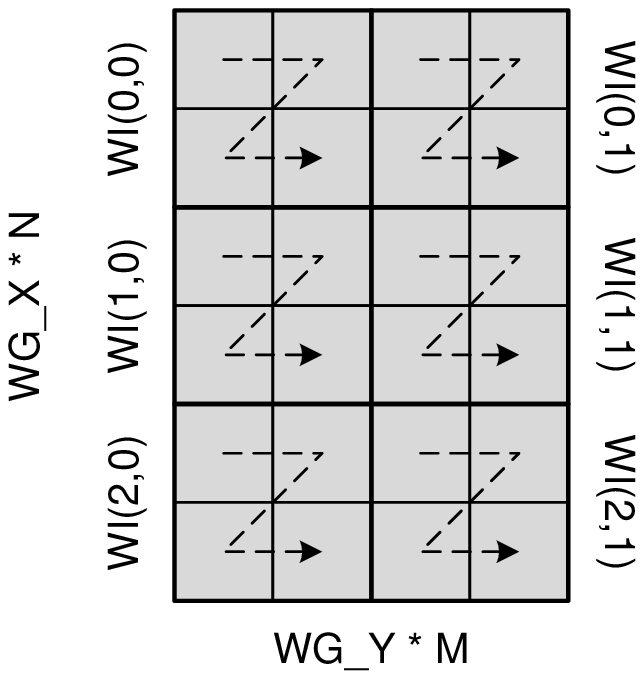}
\label{fig:hap-no-reuse-col-major}
}
\caption{Home access pattern visualization.}
\label{fig:home-access-pattern-visual}
\end{figure*}


Accesses to array \code{in} in the inner loop body are centered around a {\em
home coordinate} \code{(idx\_o, idx\_i)}, with different constant offsets in
each dimension: \code{CO\_1}, ..., \code{CO\_k}, \code{CI\_1}, ...,
\code{CI\_k} (lines 25 and 27).
The home coordinate is a linear function of the current work unit coordinate
\code{(wu\_x, wu\_y)} and the loop iterators \code{i} and \code{j}, as
specified by \code{fo} and \code{fi} at lines 23 and 24.


The 2D grid of work units is distributed across a 2D space of \wgs\ in a
blocked manner, and is further distributed across a 2D space of \wis\ in a
cyclic manner.
The geometry of array \code{out} and the launch configuration collectively
determine the number of work units each \wi\ processes (\code{NUM\_WUS\_X} and
\code{NUM\_WUS\_Y} at lines 12 and 13.
The current work unit coordinate \code{(wu\_x, wu\_y)} is computed based on
work group ID \code{(wg\_x, wg\_y)}, work item ID \code{(wi\_x, wi\_y)} and
work unit ID \code{(iter\_x, iter\_y)}, at line 15.

\begin{figure}[ht!]
\centering
\subfloat[Rectangular.]{
\includegraphics[width=0.8in]{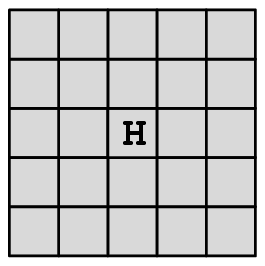}
\label{fig:neighbourhood-pattern-rect}
}
\quad
\subfloat[Diamond.]{
\includegraphics[width=0.8in]{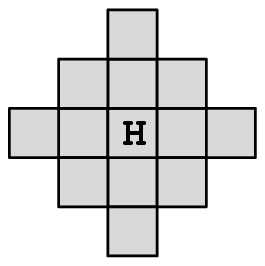}
\label{fig:neighbourhood-pattern-diamond}
}
\quad
\subfloat[Star.]{
\includegraphics[width=0.8in]{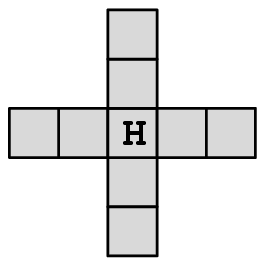}
\label{fig:neighbourhood-pattern-star}
}
\caption{Stencil patterns of target array accesses. The element of home
coordinate is labeled with ``H''.}
\label{fig:syn-kernel-neighbourhood-patterns}
\end{figure}

The kernel template provides the flexibility to vary those characteristics
that may affect the benefit of caching \code{in} data in the local memory.
First, the data access pattern to the target array \code{in} is configurable.
It is collectively defined by 1) the function tuple (\code{fo}, \code{fi})
that determines the {\em home access pattern}, and 2) the offsets \code{CO}'s
and \code{CI}'s that determine the {\em stencil pattern} (within each \wi).
We design 7 function tuples, shown in
Figure~\ref{fig:home-access-pattern-visual}, that correspond to regular access
patterns with potentially different degrees of data reuse and memory
non-coalescing.
In each diagram, an arrow indicates the order of home coordinates (of array
\code{in}) a \wi\ accesses as it goes through the iterations of loop \code{i}
and \code{j} (line 21 and 22 of Figure~\ref{fig:syn-kernel-template}).
Each arrow is associated with a label that indicates what \wis\ of a \wg\ make
such accesses.
For example, \code{WI(1,*)} refers to all \wis\ with \code{wi\_x = 1}.
Combined with the stencil pattern, this label reflects the amount of data
reuse.
The entire grey region shown in each diagram, extended with apron regions that
cover neighbouring accesses, corresponds to the region of \code{in} to be
cached in the local memory, reflecting the amount of resources consumed.
We use three common stencil patterns: rectangular, diamond and star, shown in
Figure~\ref{fig:syn-kernel-neighbourhood-patterns}.
The target array \code{in} is padded to ensure no out-of-bound accesses.


In addition to the data access pattern, the amount of computation and memory
accesses in the inner loop body and the epilogue are also configurable.
%
%
They collectively model {\em contextual} kernel computation and memory
accesses that may affect the optimization's benefit with a drop in
parallelism.
Further, they enable variation of kernel register usage.


Table~\ref{tab:syn-kernel-params} summarizes the list of 13 parameters the
kernel template provides.
Run-time parameters are shown in italics; the rest are compile-time
parameters.


\begin{table*}[ht!]
\centering
\caption{Parameters of the synthetic kernel template.}
\label{tab:syn-kernel-params}
\begin{tabular}{|p{0.5in}|p{1.8in}|p{4.2in}|}
\hline
Category & Parameter & Description\\
\hline
Global
& {\it \code{IN\_H}}, {\it \code{IN\_W}}
& Height and width of the target array \code{in}\\
\hline
\multirow{4}{0.6in}{Data Access Pattern}
& \code{HOME\_ACCESS\_PATTERN}
& One of the seven shown in Figure~\ref{fig:home-access-pattern-visual}\\
\cline{2-3}
& {\it \code{N}}, {\it \code{M}}
& Trip-counts of loop \code{i} and \code{j}\\
\cline{2-3}
& \code{STENCIL\_PATTERN}
& One of rectangular, diamond and star\\
\cline{2-3}
& \code{STENCIL\_RADIUS}
& Radius of the selected stencil pattern\\
\hline
\multirow{3}{0.6in}{Kernel Context}
& \code{NUM\_COMP\_ILB/EP}
& \# of computation in the inner loop body and epilogue\\
\cline{2-3}
& \code{NUM\_COAL\_ACCESSES\_ILB/EP}
& \# of coalesced accesses to array \code{in2} in the inner loop body and
epilogue\\
\cline{2-3}
& \code{NUM\_UNCOAL\_ACCESSES\_ILB/EP}
& \# of non-coalesced accesses to array \code{in2} in the inner loop body and
epilogue\\
\hline
\end{tabular}
\end{table*}


\subsection{Model Design}
\label{sec:model-io}

We build a machine-learning-based model that predicts kernel speedup brought
by the local memory optimization.
It takes 18 inputs (or features) and outputs a single real number that
reflects the kernel speedup.
These inputs are:


\begin{enumerate}

\item Degree of data reuse exhibited by the home access, i.e., the average
number of \wis\ in a \wg\ where this access refers to the same array element.

\item Amount of local memory used by each \wg\ for the optimization.

\item Degree of non-coalescing exhibited by the home access (in the
unoptimized kernel), i.e., the average number of memory transactions induced
by a warp.

\item Number of accesses to the target array.

\item Minimum and maximum offsets to the home coordinate of target array
accesses (2 parameters in each dimension).

\item Number of computation operations in the inner loop body and the epilogue
(2 parameters).

\item Number of contextual memory accesses (i.e., not made to the target
array) in the inner loop body and the epilogue, and whether each is coalesced
or not (4 parameters).

\item Number of registers used per thread (in the unoptimized kernel).

\item Grid size and \wg\ size (2 parameters).

\item Number of work units each \wi\ processes, equivalent to
\code{NUM\_WUS\_X * NUM\_WUS\_Y} in the synthetic kernel template.

\end{enumerate}


It is easy to extract the above features from a synthetic kernel,
because they can be directly mapped to the parameters of the kernel template.
For example, features \#1-\#5 are computed from the template parameters that
define the data access pattern.
Features \#6 and \#7 are one-to-one correspondent to the template parameters
that define the kernel context.
Currently features are extracted automatically from the synthetic kernels {\em
when they are generated}.


To extract features from a real-world application, we must first map the
kernel structure to that of the synthetic kernel template,
by identifying the boundary of a work unit (Section~\ref{sec:syn-bench}).
We currently extract features from real-world applications {\em manually}.
However, we believe that a compiler would be able to extract them automatically
once the work unit boundary is identified.


\section{Evaluation}
\label{sec:eval}

We generate a total of 9600 synthetic kernels from the template, by varying
the parameters in Table~\ref{tab:syn-kernel-params}.
We evaluate each synthetic kernel with a number of launch configurations,
resulting in a total of 5.6 million {\em kernel instances}.
We run each kernel instance with and without the local memory optimization.


We select the values of the kernel template parameters in two steps.
First, we randomly sample 100 tuples from all compile-time parameters except
\code{HOME\_ACCESS\_PATTERN}, with the resulting value distributions listed in
Table~\ref{tab:syn-kernel-comp-params}.
Second, for each tuple, we enumerate all 7 home access patterns.
For each pattern, we enumerate a set of 4 values for \code{N} and 4 values for
\code{M}, that we perceive as common.
The value set for $N$ is ${8,16,32,64}$ for home access patterns
\code{xy-reuse} and \code{x/y-reuse-row}, and ${1,2,4,8}$ for others.
The value set for $M$ is ${8,16,32,64}$ for home access patterns
\code{xy-reuse} and \code{x/y-reuse-col}, and ${1,2,4,8}$ for others.
The target array shape (\code{IN\_H} $\times$ \code{IN\_W}) is fixed at $2048
\times 2048$.


\begin{table}[ht!]
\centering
\caption{Compile-time parameter value distribution for synthetic kernels.}
\label{tab:syn-kernel-comp-params}
{\small
\begin{tabular}{|c|c|c|}
\hline
Parameter & Value Range (Average)\\
\hline
STENCIL\_PATTERN & All three\\
STENCIL\_RADIUS & $0-2$\\
\hline
NUM\_COMP\_ILB & $5-44$ (19)\\
NUM\_COMP\_EP & $1-48$ (23)\\
\hline
NUM\_COAL\_ACCESSES\_ILB & $0-13$ (3)\\
NUM\_COAL\_ACCESSES\_EP & $0-13$ (5)\\
\hline
NUM\_UNCOAL\_ACCESSES\_ILB & $0-4$ (0.8)\\
NUM\_UNCOAL\_ACCESSES\_EP & $0-4$ (0.8)\\
\hline
\end{tabular}
}
\end{table}


For each synthetic kernel, we sweep through:
1) all valid 2D grid geometries with individual dimensions restricted to
   powers of 2 and the total size no less than 512, and
2) all valid 2D \wg\ geometries with individual dimensions restricted to
   powers of 2 and the total size no more than 1024.


In addition to synthetic kernels, we also look at eight real-world benchmarks,
summarized in Table~\ref{tab:real-bench}.
For each one, we vary kernel parameters such as launch configurations and
tiling factors, resulting in multiple kernel instances.


\begin{table*}[ht!]
\centering
\caption{Real-world benchmarks.}
\label{tab:real-bench}
\begin{tabular}{|p{0.65in}|p{0.65in}|p{3.3in}|r|r|}
\hline
Benchmark & Suite & Description & LOC & \# of Kernel Instances\\
\hline
transpose & \multirow{3}{0.8in}{NVIDIA SDK} & Matrix transpose & 6 & 21\\
\cline{1-1}
\cline{3-5}
matrixMul & & Matrix multiply ($C = A \times B$) & 9 & 330\\
\cline{1-1}
\cline{3-5}
convolution & & 2D separable convolution & 10 & 600\\
\hline
MVT & Polybench & Matrix vector multiply & 9 & 120\\
\hline
SGEMM & \multirow{4}{0.8in}{Parboil} & $C = \alpha \times A \times B + \beta
\times C$ & 10 & 48\\
\cline{1-1}
\cline{3-5}
SAD & & Computes Sum-of-Absolute-Differences between pairs of image blocks;
used in motion estimation algorithm in H.264 & 94 & 517\\
\cline{1-1}
\cline{3-5}
TPACF & & Computes the angular correlation function for a data set of
astronomical bodies & 129 & 35\\
\cline{1-1}
\cline{3-5}
MRI-GRIDDING & & Computes a regular grid of data representing an MR scan by
weighted interpolation of actual acquired data points & 126 & 35\\
\hline
\end{tabular}
\end{table*}


We collect the execution time of all kernel instances (both synthetic and
real) on NVIDIA Tesla M2090 with 6GB of memory, housed in a system with an
Intel Xeon E5-2620 CPU and 64GB of memory, running CUDA 5.0 on CentOS 6.4.
We measure the execution time of the kernel only.


Figure~\ref{fig:speedup-hist-syn} showed the distribution of synthetic kernel
speedup values brought by the local memory optimization.
Figure~\ref{fig:speedup-hist-transpose}-~\ref{fig:speedup-hist-mri-gridding}
showed the distribution of real-world kernel speedup values.
We make two observations from the figures.
First, the use of local memory is not always beneficial for both synthetic and
real-world kernels.
Second, the speedup distributions have different shapes across the synthetic
and real-world kernels.
This demonstrates the need for auto-tuning the use of local memory.


\subsection{Model Training and Evaluation}
\label{sec:ml-results}

We train a model using the performance data of randomly selected 560K
synthetic kernel instances (10\% of the total).
We build the model using Random Forest (RF)~\cite{randomforest}, from Weka
3.7.10~\cite{weka}, configured with 20 trees (of unlimited depth) and 4
attributes per tree node.


We evaluate the model's accuracy by applying it to the remaining synthetic
kernel instances and all real-world kernel instances.
We use two accuracy metrics.
The first is {\em count-based accuracy}, defined as the percentage of kernel
instances where the decision of whether or not to use local memory, predicted
by the model, matches the {\em oracle} decision based on actual kernel
performance data.
Effectively this metric assigns to each kernel instance a score of 1 when the
model predicts correctly and 0 otherwise, and computes the average across all
kernel instances.
Note that, when the model mis-predicts, the metric does not take into account
the performance loss it incurs.
Hence we introduce the second metric, {\em penalty-weighted accuracy}, which
extends the first metric by assigning a score equal to the performance ratio
(a value between 0 and 1), instead of 0, when the model mis-predicts.
Effectively this metric measures the percentage of kernel performance
achieved using the model-predicted decision, over that achieved by the oracle
decision, averaged across all kernel instances.

\begin{figure}[ht!]
\centering
\includegraphics[width=3.3in]{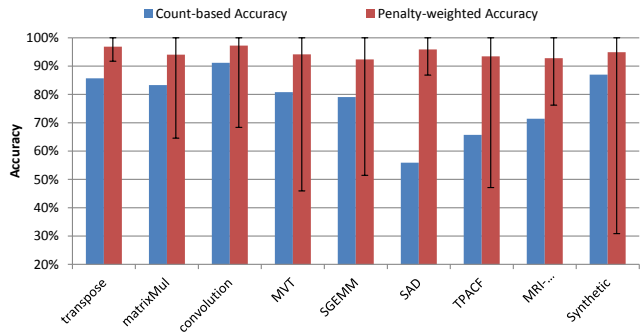}
\label{fig:model-accuracy-syn-training}
\caption{Accuracy of machine-learning-based models.}
%
%
\end{figure}

Figure~\ref{fig:model-accuracy-syn-training} shows the model accuracy, with
both metrics, for synthetic and real-world kernels.
For penalty-weighted accuracy, we also show the range (min-max) of
per-kernel-instance scores using error bars.
Overall, the trained model achieves 86\% count-based accuracy and nearly 95\%
penalty-weighted accuracy on the remaining synthetic kernel instances.
However, a 30\% minimum score indicates that the model does mis-predict on a
small percentage of kernel instances with high performance penalty.
For real-world benchmarks, the model is able to achieve nearly 95\%
penalty-weighted accuracy, although the count-based accuracy drops noticeably
for SAD, TPACF and MRI-GRIDDING.
This shows that the model trained with a large number of synthetic kernels can
achieves high accuracy consistently across a variety of kernels.


\section{Related Work}
\label{sec:related-work}

There has been growing interest in the use of machine learning to auto-tune
the performance of GPU
applications~\cite{liu-ipdps09,bergstra-inpar12,grewe-cgo13,magni-gpgpu13,magni-pact14}.
For example, Magni et al.~\cite{magni-pact14} explore the use of neural
network for auto-tuning thread coarsening, and Grewe et al.~\cite{grewe-cgo13}
use a decision tree to decide if an OpenCL kernel should be executed on the
CPU or the GPU.
In contrast, we focus on auto-tuning the use of local memory.


There is work that explored auto-tuning of the use of local memory, but
focused on the use of analytical modeling and empirical
search~\cite{ryoo-jpdc08}.
In contrast, we build machine learning models, which have the potential to be
more accurate than analytical approaches.


Finally, there is a large body of work that treats auto-tuning for platforms
other than GPUs, including multi-cores~\cite{ganapathi-hotpar09,wang-pact10}
and single-core
processors~\cite{stephenson-pldi03,cavazos-pldi04,furin-ijpp11}.
In contrast, we focus on GPUs.


\section{Conclusion and Future Work}
\label{sec:conclusion}

In this paper we described a machine learning model for use in automatic
performance tuning of local memory usage on GPUs.
We have shown that the optimization is not always beneficial, depending on
application characteristics.
We train a Random Forest model with a large number of synthetic benchmarks and
predict whether local memory should be used or not for a single array access
in both synthetic and real benchmarks.
We have shown that the penalty-weighted prediction accuracy is nearly 95\%.


This work can be extended in various directions.
First, the use of this model in practice demands a compiler framework that
automatically applies the local memory optimization and extracts kernel
features.
Second, the model can be extended to predict the {\em usage} of local memory
when multiple arrays compete for local memory resources.
Third, the prediction accuracy can be evaluated for a larger set of real-world
benchmarks.
Fourth, the quality of the machine learning model using synthetic benchmarks,
as opposed to real benchmarks, can be evaluated.
Finally, other machine learning models (e.g., Support Vector Machines) can be
evaluated.



\acks
This work was supported by grants from the Natural Sciences and Engineering
Research Council of Canada (NSERC) and Qualcomm, Inc., and by an NVIDIA
Graduate Fellowship.



\bibliographystyle{abbrvnat}



\end{document}